\documentclass[prl,twocolumn,showpacs,superscriptaddress]{revtex4}

\usepackage{graphicx}
\usepackage{dcolumn}
\usepackage{bm}
\usepackage{amsmath}
\usepackage{latexsym}

\begin{document}
\title{A spatial network explanation for a hierarchy of urban power laws}
\author{Claes Andersson}
\email{claes@santafe.edu} \affiliation{Department of Physical
Resource Theory, Chalmers University of Technology, 412 96
G\"oteborg, Sweden}

\author{Alexander Hellervik}
\email{f98alhe@dd.chalmers.se} \affiliation{Department of Physical
Resource Theory, Chalmers University of Technology, 412 96
G\"oteborg, Sweden}

\author{Kristian Lindgren}
\email{frtkl@fy.chalmers.se} \affiliation{Department of Physical
Resource Theory, Chalmers University of Technology, 412 96
G\"oteborg, Sweden}

\date{\today}

\begin{abstract}
In this Letter we report an application of spatial complex
networks to simulate the evolution of the spatial economic system.
The model depicts the spatial trade network with land areas as
nodes and trade relations between activities as connections. The
network grows by addition of new pairs of activities using simple
market economic assumptions. The definition of nodes and
connections allows us to compare model and reality on the level of
market land values. The model provides an explanation to several
empirically observed statistical phenomena in spatial economics.
Furthermore, market demand is an important driving force for most
other spatially extended systems of human origin. Thus, a model
that simulates the geographic distribution of economic activity
may shed light on a range of other systems that have been
investigated separately using complex networks.
\end{abstract}

\pacs{61.43.Hv, 89.75.Da, 89.75.Hc, 89.65.Gh}

\maketitle

Standard models of urban economics have difficulties describing
economic growth and as a consequence they cannot reproduce
important empirical observations, maybe most notably Zipf's
Law\cite{Zipf_1949,Ioannides_Overman_2002,Fujita_Krugman_Venables_1999}.
One of the basic tenets of market economics is that economic
growth arises as a result of mutually beneficial trade between
specialized agents. We use this mechanism as the basis of an
abstract network representation of economic activity and trade on
a surface.

Casting the urban economy as a complex network puts a range of new
tools at our disposal for analyzing this important
system\cite{Albert_Barabasi_2002,Dorogovtsev_Mendes_2002,Newman_2003}.
The benefit goes beyond direct observations about land uses and
land values since many other growing spatial systems of human
origin grow as a direct response to demand from within the
economy. Because of this, a model that simulates the geographic
distribution of economic activity can also shed light on the
workings of many other social, technical and economical systems --
many of which has been studies as complex networks in their own
right such as the Internet, phone call graphs and human sexual
contacts
\cite{Albert_Jeong_Barabasi_1999,Pastor-Satorras_Vazquez_Vespignani_2001,Aiello_Chung_Lu_2000,Liljeros_Edling_Amaral_Stanley_Aberg_2001,Albert_Barabasi_2002,Dorogovtsev_Mendes_2002,Newman_2002a}.
It is quite conceivable that reported power law distributions in
these and other such systems may stem more or less directly from
near-optimal distribution of services to meet demand within the
geographic distribution of the urban system.

We take fixed-size non-overlapping land areas to be network nodes,
and the trade relations between them to be connections, in the
network model. The activity within a node is defined as the number
of connections connecting it to the rest of the network. We hereby
obtain an abstract representation that can capture heterogeneity
in producer specialization and spatiality at the same time. This
allows us to introduce a basic economic unit, the Economically
Correlated Pair (ECP), as an abstract expected future stream of
trade between two activities. Per definition, an ECP has unit
economic value to both its end-points -- a reasonable
simplification given the very large number of trade relations in
an economy.

The simulations we use for the results in this Letter start from a
lattice of undeveloped land areas and proceeds to simulate the
allocation of new urban activities and their trade relations. We
compare statistics of simulated node degrees with empirically
observed land values. The comparison assumes a linear relationship
between the amount of money generated and the value of land. This
relationship follows from our definition of ECPs and from i)
market pricing of goods and services and ii) the connection
between trade benefits and land value.

 i) Market pricing of commodities provides an adaptive measure
that allows us to compare the activities that generate them.
Hence, on average, an edge contributes identically to the value of
both nodes to which it connects. This contribution is exactly our
definition of activity, which implies that the degree of a node is
proportional to its benefits due to trade.

ii) This connection consists of two proportionalities. For a node
$i$ we have
\begin{equation}
v_i \propto r_i \propto x_i,
\end{equation}
where $v_i$ is the value of the corresponding land area, $r_i$ is
the bid-rent\cite{Alonso_1964}, and $x_i$ is the total trade
benefits as outlined above. Capitalizing periodic rent income from
the site $i$ gives land value $v_i=\frac{r_i}{i}$ where $i$ is the
interest rate\cite{OSullivan_2002}. The second proportionality is
a weak form of the leftover principle from urban economics, which
states that, in a competitive land market, rent equals the amount
of money left after all expenses (except rent) are paid. This
amount of money equals the sum of all trade benefits at the site.
For our results it is sufficient that, on average, a certain
proportion of each new unit of trade benefit goes to the
landowner.

We formulate the network model by dividing a geographic area into
$N$ non-overlapping fixed-size cells $\{1,2,\ldots, N\}$ and
taking these to be network nodes. The undirected connections
between nodes are trade streams (ECPs) between activities in the
cells, these can also connect a cell to itself and two nodes can
be connected by any number of mutual ECPs. The amount of profit
(before rent), generated in a node $i$, is per the definition of
ECPs equal to $mx_i$, where $m$ is the value of an ECP and $x_i$
is the degree.

Network evolution is the process by which a network changes over
time. New activity occurs in the form of additions of ECPs to the
system. The growth mechanisms include balanced node addition and
preferential attachment which gives a strong similarity to the
Barab\'asi-Albert model, and a scale-free node distribution can be
expected\cite{Barabasi_Albert_1999}.

The network is initialized by the pairwise connection of $n_0$
(even) spatially uniformly distributed nodes. Addition of ECPs are
separated into primary and secondary growth, corresponding to the
placing of first and second edge end-points in the network. The
placement of an end-point is done either preferentially
(per-activity) or uniformly (per-node); corresponding to \emph{in
situ} expansion or establishment of new activity, respectively.
Expansion of node activity can correspond to existing activities
becoming more space efficient as well as them being out-bid and
replaced by more efficient activities.

Preferential growth corresponds to trade with an existing
activity. Activities are considered to be average activities and
thus they all interact to the same extent with each other. If we
assume constant fractions $q_1$ and $q_2$ for uniform and
preferential growth respectively, with $q_1+q_2=1$, the
probability of selecting a node $i$ preferentially in primary
growth is
\begin{equation}
\Pi_i^{1,pref}=q_2\frac{x_i}{\sum_jx_j}. \label{eq:BA1}
\end{equation}
Uniform growth corresponds to average activities selecting sites
in a competitive land market. The probability of selecting a node
$i$ uniformly as a primary effect is
\begin{equation}
\Pi^{1,uni}_i=q_1\frac{\delta^{(D)}_i + b\delta^{(P)}_i +
b\epsilon\frac{n_t^{(P)}}{n_t^{(E)}}\delta^{(E)}_i
}{n_t^{(D)}+b(1+\epsilon)n_t^{(P)}}\label{eq:UNI1},
\end{equation}
where $n^{(D)}_t$ is the number of developed nodes at time $t$ and
$\delta^{(D)}_j=1$ if node $j$ is developed and $\delta^{(D)}_j=0$
otherwise. The meanings of $n^{(P)}_t$, $n^{(E)}_t$,
$\delta^{(P)}_j$ and $\delta^{(E)}_j$ are analogous to $n^{(D)}_t$
and $\delta^{(D)}_j$, with $P$ referring to perimeter nodes and
$E$ referring to external nodes. Perimeter nodes are all
undeveloped nodes that are adjacent to a developed node and
external nodes are undeveloped nodes without any developed
neighbors. The parameter $b$ controls the fraction of perimeter
nodes that are open for development and $\epsilon$ controls how
many external nodes that are open for development compared to the
number of perimeter nodes. The interpretation of this is that
external growth is identical to perimeter growth but takes place
on ambient infrastructure not explicitly represented in the model.
Infrastructure tends to grow as the urban system grows which makes
it reasonable to assume that $\epsilon$ is constant over time.

The probability of secondary preferential growth at site $i$ as a
consequence of primary growth at site $j$ is
\begin{equation}
\Pi_i^{2,pref}=q_2\frac{D_{ij}x_i}{\sum_kD_{kj}x_k},
\label{eq:interact}
\end{equation}
and for secondary uniform growth, it is
\begin{equation}
\Pi^{2,uni}_i=q_1\frac{D_{ij}\left(\delta^{(D)}_i +
b\delta^{(P)}_i
+b\epsilon\frac{n_t^{(P)}}{n_t^{(E)}}\delta^{(E)}_i\right)}{\sum_kD_{kj}\left(\delta^{(D)}_k
+ b\delta^{(P)}_k +
b\epsilon\frac{n_t^{(P)}}{n_t^{(E)}}\delta^{(E)}_k\right)
}\label{eq:UNI2}.
\end{equation}
$D_{ij}$ is a matrix of site-to-site interaction strengths, which
decay with transportation costs. We have used $D_{ij}=(1+c
d(i,j))^{-\alpha}$, where $d(i,j)$ is the Euclidean distance
between sites $i$ and $j$. The non-negative parameters $c$ and
$\alpha$ controls the impact of spatiality.

\begin{figure*}
\includegraphics[width=18.0cm, height=10.0cm]{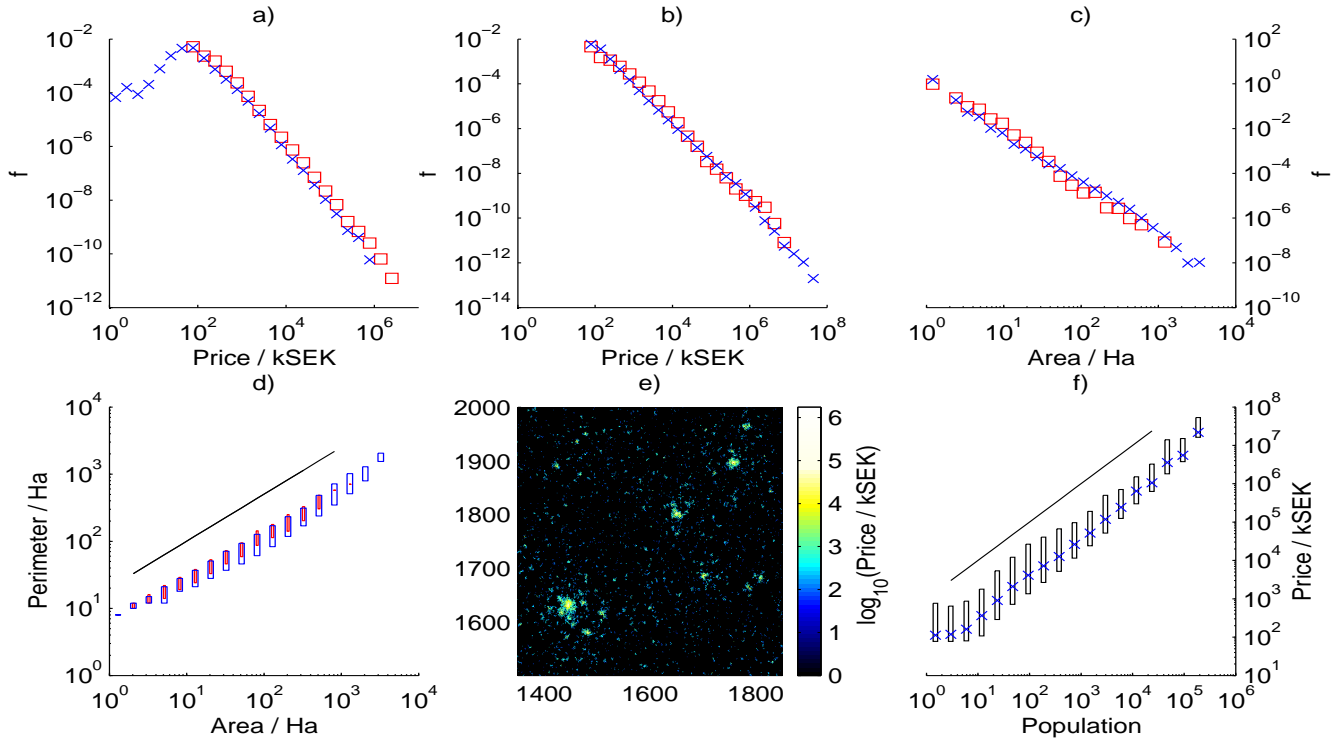}
\caption{\label{panel} Diagrams (a), (b), and (c) show
double-logarithmic histograms with exponentially binned empirical
($\times$) and simulated ($\Box$) observables: (a) - land value
per cell, (b) - aggregated cluster land value, (c) - cluster area.
The lower part of the empirical distribution in (a) corresponds to
land values below our urban land-use threshold $m=75$kSEK/Ha, and
is not described by our model. The exponents of the power law
distributions in (a), (b) and (c) are roughly $2.1$, $1.8$ and
$2.3$ respective. In (d) is shown empirical (broad boxes) and
simulated (thin boxes) results for cluster area plotted against
exponentially binned cluster perimeters. The vertical interval of
the boxes contains 90\% of the perimeters in the bin. The
reference line has a slope of $0.7$. Part of the simulated
configuration used for the statistics shown in the other figures
is shown in (e). Diagram (f) shows cluster populations plotted
against exponentially binned cluster prices. The vertical interval
of the boxes contain 90\% of the cluster prices in the
corresponding bin and the crosses indicate the medians of the
prices in the bins. The reference line has a slope of 1, which
indicates a linear relationship. The comparison was carried out by
first identifying geographical clusters of high land values and
then comparing their accumulated population content with their
accumulated value. }
\end{figure*}

To simplify the analysis of the model it is useful to assume that
development of new land (addition of an active node) takes place
at a constant rate $q_A$ (compared to other types of growth, not
as a function of physical time). To motivate the assumption, let
us consider the growth of developed clusters and their perimeters.
It is true in simulations of the model, and it can be empirically
verified (see Figures \ref{panel}c and \ref{panel}d), that the
cluster area distribution is close to a simple power-law with
density function $f(A)\sim A^{-\beta}$, and that the relation
between cluster perimeter $P$ and cluster area $A$ has the form
$P\sim A^{\lambda}$, with $\lambda<1$. From this we observe that
for the entire system of clusters we have
\begin{equation}
\frac{n_t^{(P)}}{n_t^{(D)}}\sim \frac{\int^{\infty}_{1}
A^{-\beta}A^{\lambda} d A}{\int^{\infty}_{1} A^{-\beta}A d
A}=\frac{\beta-2}{\beta-\lambda-1},\label{eq:PerArea}
\end{equation}
assuming that $\beta>2$ and $\lambda<1$. We define $q_1'$ as the
fraction of activity increments that occur uniformly on developed
nodes, and Eq. (\ref{eq:UNI1}) gives
\begin{equation}
q_1'=\sum_i\delta^{(D)}_i\Pi^{1,uni}_i=q_1\left(1+b(1+\epsilon)\frac{n_t^{(P)}}{n_t^{(D)}}\right)^{-1},
\end{equation}
which, because of Eq. (\ref{eq:PerArea}), is approximately
constant. This means that the rate of node activation,
$q_A=1-q_1'-q_2$, also can be considered constant.

If we assume that the spatial bias between sites is
small\cite{Andersson_Hellervik_Lindgren_2003}, the time evolution
of expected activity on a developed site $i$ follows the equation
\begin{equation}
x_i(t+1)=x_i(t)+2q_1'\frac{1}{n^{(D)}_t}+2q_2\frac{x_i(t)}{\sum_jx_j(t)}\label{eq:meanBA},
\end{equation}
which is solved by the continuous-time method introduced by
Barab\'asi et al \cite{Albert_Barabasi_Jeong_1999}. After
sufficiently long time the degree distribution approaches the form
$\label{eq:power} P[x_i=x]\sim(x+B)^{-\gamma}$, with
\begin{equation}
B=\frac{q_1'}{q_2(1-q_1'-q_2)}
\end{equation}
and
\begin{equation}
\gamma=1+\frac{1}{q_2}\label{eq:exponent}.
\end{equation}

Most of the parameters in the model can be readily estimated from
empirical data. If $n^{(C)}_t$ is the number of clusters, we have
$\epsilon=n_t^{(C)}/n_t^{(D)}$. The exponent $\gamma$ in the
distribution of activity gives information about the size of $q_2$
(and $q_1$) via Eq. (\ref{eq:exponent}). The perimeter parameter
$b$ can be determined by $q_A=\frac{n_t^{(D)}}{\sum_ix_i(t)}$ and
$b=q_A\left((q_1-q_A)(1+\epsilon)\frac{n_t^{(P)}}{n_t^{(D)}}\right)^{-1}$.

In principle, these parameters are not empirical in nature, they
are aggregated measures of fundamental properties of the current
economic system (and region) of interest. The spatial parameters
$c$ and $\alpha$ reflect the statistics of transport
characteristics of the economic configuration. They are not as
easily estimated from data as the other parameters but results
seem robust to their exact values and the functional form of
$D_{ij}$.

The data used for all empirical results is based on a database
delivered by Sweden Statistics that covers estimations of the
market value of all land in Sweden (2.9 million data points).
Cluster measurements was obtained by identifying clusters using
land value as the threshold parameter (75 kSEK/Ha) and then
comparing accumulated price and population within the obtained
areas. To identify clusters in simulated and empirical data we use
a computer program that masks away all data points below a
threshold value and then treats all contiguous (8-cell
neighborhood) areas as clusters.

Once the frequency distribution of land value per unit area is
correct, the challenge in realizing the cluster level is to
faithfully capture the spatial sorting of the node degrees. Using
the model presented in this Letter we obtain excellent agreement
between simulated and empirical statistics for a range of
non-trivial higher-order structures in successive orders of upward
causation (cells to clusters), see Figures (\ref{panel}b) through
(\ref{panel}b). These results are not trivial consequences of the
land price distribution; it is perfectly possible to arrange the
developed cells into any system of clusters. The same is true for
the relation between cluster area and perimeter. Also, in Figure
(\ref{panel}f) we demonstrate that geographical distribution of
population and land value can be observed interchangeably. This is
a likely example of how an earlier studied spatially growing
system of human origin likely is just a response to the growth of
the economic network. The growth of the latter is defined by
activity inter-dependency and transportation costs -- a more
fundamental candidate for a generating process than more vague
postulations about why and when people
re-locate\cite{Marsili_Zhang_1998}.

All figures reflect data from the same Monte Carlo simulation run.
Figure (\ref{panel}e) shows a part of the spatial configuration
from this run. The parameters for the reported simulation,
$m=75$kSEK ($10$SEK$\approx 1$USD), $b=0.15$ and $\epsilon=0.25$
have been estimated from empirical data. By studying the exponent
in the empirical distribution of land values, an initial estimate
of $q_1=0.1$ was obtained. For the reported results $q_1$ was
adjusted to $0.2$. The spatial parameters were $c=0.1$ and
$\alpha=2$. Investigation of sensitivity shows that exponents and
proportions change slowly and smoothly with all parameters. A
square grid of $2000\times2000$ cells was used and the number of
iterations was $10^6$.

Macroscopic models, such as Simon's Model and later derivations,
are based on the rather unrealistic assumption that individual
cities would react in unison to external
events\cite{Simon_1955,Gabaix_1999,Schweitzer_Steinbrink_1998}.
Although this may in some sense be
true\cite{Ioannides_Overman_2002}, such models hold limited
explanatory and predictive power since they can not be mapped to a
realistic situation where the internal structure of cities is
important or when clusters interfere and coagulate. By making a
similar assumption for small fixes-size land areas our
representation remains valid regardless of the dynamics of
higher-order structures such as cities.

The relation between pattern (probability distributions etc.) and
process (microdynamics) is not one-to-one. Rather, for each
observed macroscopic pattern there is likely a whole class of
processes capable of generating it. Apart from arguing for the
network model ontology on the micro level there is a number of
implications that remain to test against empirical data. Among
these are the time evolution of land values, explicit measurement
of the trade network and making studies similar to the one
presented here for other countries.


We thank Anders Hagson and Jonas Tornberg, City and Mobility,
Chalmers University of Technology, for providing preprocessed
empirical data. We also thank Martin Nilsson and Steen Rasmussen
for valuable discussions and input.
\bibliographystyle{unsrt}
\bibliography{references}

\end{document}